\documentclass[double]{article}
\usepackage{times}
\usepackage{algorithm}
\usepackage{algorithmic}
\usepackage{wrapfig}
\usepackage{verbatim}
\usepackage{amsmath}
\usepackage{graphicx}
\usepackage{subcaption}
\usepackage[active]{srcltx}
\usepackage{color}
\usepackage{lineno}
\usepackage{authblk}

\usepackage{epsfig,graphicx,amsfonts}
\usepackage{amsmath,amsthm,amssymb}
\usepackage{xcolor}

\usepackage{adjustbox}
\usepackage{multirow}
\usepackage{setspace}
\usepackage{hyperref}
\usepackage{comment}

\long\def\comment #1\commentend{}

\begin{document}

\title{\Large Cost-optimal Seeding Strategy During a Botanical Pandemic in Domesticated Fields}
\author{Teddy Lazebnik$^{1,2,*}$ \\
$^{1}$Department of Mathematics, Ariel University, 4070000, Ariel, Israel\\
$^{2}$Department of Cancer Biology, Cancer Institute, University College London, WC1E 6BT, London, UK\\
 $^{*}$Corresponding author: lazebnik.teddy@gmail.com 
}

\maketitle 

\linespread{1.5}

\begin{abstract}
\noindent
Botanical pandemics cause enormous economic damage and food shortages around the globe. However, since botanical pandemics are here to stay in the short-medium term, domesticated field owners can strategically seed their fields to optimize each session's economic profit. In this work, we propose a novel epidemiological-economic mathematical model that describes the economic profit from a field of plants during a botanical pandemic. We describe the epidemiological dynamics using a spatio-temporal extended Susceptible-Infected-Recovered epidemiological model with a non-linear output economic model. We provide an algorithm to obtain an optimal grid-formed seeding strategy to maximize economic profit, given field and pathogen properties. We show that the recovery and basic infection rates have a similar economic influence. Unintuitively, we show that a larger farm does not promise higher economic profit. Our results demonstrate a significant benefit of using the proposed seeding strategy and shed more light on the dynamics of the botanical pandemic. \\

\end{abstract}

\maketitle \thispagestyle{empty}


\pagestyle{myheadings} \markboth{Draft:  \today}{Draft:  \today}
\setcounter{page}{1}

\vspace{1cm}
\noindent
\textbf{Lead paragraph:} Botanical pandemics are common over history, causing both food shortages as well as enormous economic damage. In this study, we proposed a seed strategy for cultivated fields using a novel spatiotemporal epidemiological-economic model that takes into consideration pandemic spread dynamics and its influence on the field's economic output. In this highly chaotic environment, we show an algorithm that proposes the optimal seeding configuration given information about a specific pathogen to maximize economic output. The proposed model can be adopted to help farmers be more prepared to next botanical pandemics. 

\onehalfspacing
\section{Introduction}
\label{sec:introduction}

For thousands of years, humankind based its food supply on agriculture \cite{intro}. Multiple historical records show large-size agricultural pandemics that cause heavy losses for farmers and society \cite{intro_2}. In particular, multiple plant virus disease pandemics and major botanical pandemics occurred worldwide in the last century \cite{bp_1,bp_2,bp_3,s_reviewer_1,s_reviewer_2}. Overall, the number of agricultural crop epidemics exhibits a monotonic increase while recently reaching an asymptotic rate \cite{related_1}. 

In recent times where agriculture is part of the multi-sectoral economy, an agricultural pandemic has a two-folded impact on society: food shortages and direct economic losses to the agriculture sector \cite{intro_3,pandemic_economy}. For instance, during 2014 alone, virus disease pandemics were estimated to have a global economic impact of at least 30 billion dollars \cite{intro_money}. Moreover, as farmers and companies aim to increase their profit from the same land and to cope with growing demand, they adopted multiple methods such as the cultivation of annual crops as monocultures and plant breeding \cite{intro_economy_1,intro_economy_2}. As a direct result, the plants' immunity is often harmed due to these practices, which provide an underlying basic ingredient of instability \cite{intro_economy_3}. In repercussion, severe virus disease epidemics became regularly recurring features in many herbaceous crops \cite{intro_economy_2}. Hence, it is of interest to farmers and policymakers to control such pandemics, minimizing their negative impacts.

Mathematical and computational models are key tools for understanding pandemic spread and designing intervention policies that help control a pandemic's spread. In particular, coupled ordinary and partial differential equations, as well as simpler growth-curve equations, are especially useful deterministic models for representing plant disease development in fields \cite{intro_3,intro_related_1,intro_related_3,intro_related_4}. The authors of \cite{related_1} used the Susceptible-Infected-Removed (SIR) model, originally proposed by \cite{first_sir_paper} to describe pandemics in humans, to examine the effects of different models for the effect of host responses to a load of infection on the production of susceptible tissue. The authors tested their model on the stem canker disease of potatoes caused by the soil-borne fungus (Rhizoctonia Solani), showing a promising prediction capability on historical data. The authors of \cite{related_2} utilized a Healthy-Latently-Infected-Diseased (HLD) model for tomato bacterial canker (TBC) caused by the pathogenic plant bacteria Clavibacter Michiganensis Subsp. Michiganensis (Cmm). They assumed the infection was transferred to healthy plants through contaminated scissors used to cut symptomless infected plants and fitted the model on a dedicated experiment. Their results show that the model can fairly predict the number of diseased plants over time. The authors of \cite{related_3} extended the SIR model, proposing an SIRX model that incorporates two sources of infection, with primary infection arising from 'free-living' inoculum and secondary infection occurring by transmission from infected to susceptible hosts. Their model focused on soil-borne plant diseases caused by various fungal and bacterial pathogens in crops. The authors analyzed the sensitivity of various epidemiological and botanical properties, showing that the infection rate is susceptible to most of them.

This paper focuses on the grid-based seeding strategy during a botanical pandemic from an economic perspective. Namely, the novelty of the proposed work is the formalization and analysis of a two-parametric seeding strategy for a field experiencing a botanical pandemic. To accomplish this objective, we developed a spatio-temporal stochastic SIR model with an economic dynamic where the plants are organized on a two-dimensional grid, seeded at the beginning of a season, and harvested and sold at the end of the season.

The remainder of this paper is organized as follows. Section \ref{sec:model} formally introduces the proposed model and formalizes the objective. Next, section \ref{sec:worse_case} describes an algorithm to obtain the optimal seeding strategy for the worst-case pandemic, given initial conditions. Afterward, section \ref{sec:results} presents \textit{in silico} experiment of the proposed model, analyzing the sensitivity of the pandemic spread and economic profit to the pathogen's properties and the optimal seeding strategy for different scenarios. Lastly, in section \ref{sec:discussion} we discuss the results and suggest possible future work.

\section{Model definition}
\label{sec:model}
\begin{figure}[!ht]
    \centering
    \includegraphics[width=0.99\textwidth]{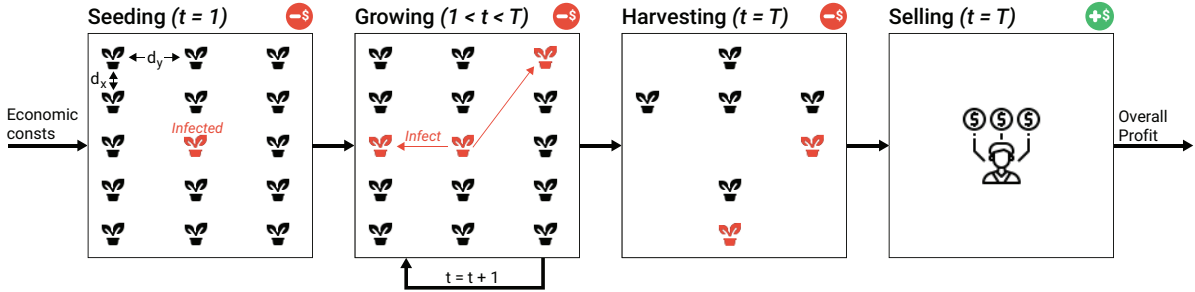}
    \caption{A schematic view of the model's spatio-temporal dynamics. At the beginning (\(t=1\), the field is seeded and some plants are infected. Next, for a duration, \(T < \infty\), the plants grow and infect each other. After \(T\) steps in time, the plants are harvested and the non-infected plants are sold for profit.}
    \label{fig:model_design}
\end{figure}

The proposed model \(\mathbb{M}\), is defined as a tuple of two interconnected components \(\mathbb{M} := (\mathbb{P}, \mathbb{E}\)) where \(\mathbb{P}\) is the epidemiological component responsible for capturing the spatio-temporal pandemic spread of a pathogen in a population of plants, and \(\mathbb{E}\) is the economy component responsible for capturing the economic profit the field's owner obtain over time. Below, a formal description of the two components and their interactions is provided. Moreover, a description of the model's implementation as a computer simulation is also provided. A schematic view of the model's dynamics over time (\(t \in [1, \dots, T\)) is presented in Fig. \ref{fig:model_design}.

\subsection{The epidemiological component}
Let us define a sub-model (i.e., component) that takes into consideration a population of plants, \(P\) \((|P| := N \in \mathbb{N})\), that is allocated in a finite and rectangular-shaped field \(F := (W, H) \subset \mathbb{R}^2\). The model treats both space and time as continuous variables. Importantly, in Section \ref{sec:simulation}, during the implementation of this model as a computer simulation, we would discrete both space and time \cite{teddy_ariel,abs_1,abs_2,abs_3}). Interactions are local and stochastic. It is further assumed that \(F\) is homogeneous and isotropic \cite{spatial}. The plant population, \(P\), is allocated to the field \(F\) such that it creates rows and columns with a distance \(d_x\) and \(d_y\) between them, respectively. 

Each plant in the population, \(p \in P\), belongs to one of three epidemiological groups: Susceptible (\(S\)), Infected (\(I\)), or Removed (\(R\)) such that \(N := S + I + R\). Plants in the first group have no immunity and are susceptible to infection. When a plant in the susceptible group (\(S\)) is exposed to the pathogen, the plant is transferred to the infected group (\(I\)), at an average rate \(\beta\). The plant stays in the infected group for \(\gamma\) steps in time, after which the plant is transferred to the removed group (\(R\)). Removed plants do not practice in any epidemiological nor economic dynamics. We define the infection rate, \(\beta\), to be \textit{distance-depended}. Particularly, the rate at which plant \(p_i\) is infecting plant \(p_j\) is defined to be:
\begin{equation}
\beta(p_i, p_j) := \frac{\beta_0}{\|p_i - p_j\|},
\label{eq:infection_rate}
\end{equation}
where \(\beta_0 \in [0, 1]\) is the basic infection probability of the pathogen and \(\|p_i - p_j\|\) stands for the Euclidean distance between the plants \(p_i\) and \(p_j\). Thus, the spatio-temporal epidemiological dynamics obey the following system of coupled ordinary differential equations:
\begin{equation}
    \begin{array}{l}
        \frac{dS(t)}{dt} = - \beta S(t) I(t), \\\\ \frac{dI(t)}{dt} = \beta S(t) I(t) - \gamma I(t), \\\\  \frac{dR(t)}{dt} = \gamma I(t).
    \end{array}
\end{equation}

\subsection{The economic component}
Continuing the definition of the epidemiological component, let us define the economic sub-model of the same population of plants over time. The economic component considers the cost of seeding, growing, and harvesting plants and the profit one obtains from selling them. Formally, at the beginning of the session, one is required to seed the plants. This process is associated with a cost for each plant (\(0 \leq a_1^s \in \mathbb{R}\)) and some fixed overhead cost (\(0 \leq a_2^s \in \mathbb{R}\)). However, the cost per plant is growing in a logarithmic manner to the size of the plant population \cite{price_1,price_2}. Thus, for a plant population of size \(N\) the seeding cost is \(O_{seeding}(N) = a_1^s ln(N) + a_2^s \). In the same manner, the cost of growing the plants in a single step in time corresponds to some fixed overhead and a cost per plant, increasing in a logarithmic manner: \(O_{growing}(N) = a_1^g ln(N) + a_2^g\). Finally, the cost of harvesting a plant population follows the same dynamics: \(O_{harvesting}(N) = a_1^h ln(N) + a_2^h\). After harvesting, the plants can be sold. The plants has a fixed price, \(0 < \psi_1 \in \mathbb{R}\), while the overall sales size results in a reduced value, \(0 < \psi_2 \in \mathbb{R}\), in a logarithmic manner to the size of the plant population \cite{economic_1,economic_2,economic_3}: \(O_{sell}(N) = \psi_1 N - \psi_2 ln(N)\). Therefore, the economic output of the field at time \(t \in [1, \dots, T]\) is as follows: 
\begin{equation}
O(N_t) := \begin{cases}
-O_{seeding}(N_t) - O_{growing}(N_t) & t = 1 \\
-O_{growing}(N_t) & 1 < t < T \\
O_{sell}(N_t) - O_{harvesting}(N_t) & t = T
\end{cases}
\label{eq:economy},
\end{equation}
such that \(O(N_1) = 0\) and \(N_t := N - R(t)\).

Importantly, as plants require some minimal area for seeding and growing, there exists a minimal seeding distance, \(\Delta \in \mathbb{R}^+\), to allow plants to grow. A seeding with a distance smaller than \(\Delta\) will result in the early death of the plant and no profit, as a result.

\subsection{Computer simulation}
\label{sec:simulation}
In order to simulate the model, we used an agent-based simulation approach \cite{agent_based_1,agent_based_2,teddy_pandemic_management}. To this end, we assume a discrete version of the proposed model. The plants in the population interacts in rounds \(t \in [1, \dots, T]\), where \(T<\infty\). Each plant in the population, \(p_j := (x_j, y_j, \xi_{i,j}) \in \mathbb{P}\), is represented by a timed finite state machine \cite{fsm} such that, at round \(i\), \(\xi_{i,j}\) denotes the plant's epidemiological status \(p_j \in \{S, I, R\} \) and \((x_j, y_j)\) denoted the plant's location in the field (\(F\)). At the first round (\(t=1\)), the population (\(\mathbb{P}\)) is allocated to \(F\) given the values \(d_x\) and \(d_y\). In practice, during the seeding or immediately after the plants are susceptible, and only a subset is infected by a pathogen. At the beginning of the pandemic (\(t = 1\)), the seeding cost is computed. Then, at each round \(1 \leq t < T\), the pathogen spreads between the plants, and the growing cost is computed. After \(T\) rounds, the economic profit from the field, \(E\), is computed after taking the harvesting cost and profits from selling into account. 

\section{Optimal seeding strategy}
\label{sec:worse_case}
In the case of a pandemic where the pathogen properties and the economic costs are known, one can aim to optimize its profit from a given field by strategically seeding the plant population. On the one hand, seeding the plants as close to each other as possible would lead to a maximum profit, assuming the profit from selling the entire plant population is higher than the overall cost. However, on the other hand, it would increase the average infection rate (see Eq. \ref{eq:infection_rate}) and result in more removed plants that cause economic loss. 

Hence, one can formalize the above motivation as follows. Given the proposed model, one wishes to maximize the economic profit of the field during a single session (i.e., \(t \in [1, T]\)) while controlling the seeding strategy. Formally, the optimization task takes the form:
\begin{equation}
\max_{d_x, d_y \in \mathbb{R}} \sum_{t=1}^{T} O(N_t),
\label{eq:optimization}
\end{equation}
where \(d_x, d_y > 0\). Furthermore, it is assumed assume that \(k\) plants are exposed to a pathogen immediately after seeding, setting the initial condition to be \(S(0) = N - k, I(0) = k, R(0) = 0\).

Now, the locations of the initially infected plants in the field play a critical role in the pandemic spread rate, as one can see from Eq. (\ref{eq:infection_rate}). Thus, it is possible to provide a worst-case scenario as a boundary condition by choosing the \(k\) plants that the maximum distance between at least one of them to any other plant is minimal (i.e., the \textit{metric k-center} task) \cite{k_center}. Let us denote this distance by \(0 < r \in \mathbb{R}\) such that \(r := \sqrt{d_x^2 + d_y^2}\). 

Assuming the worst case scenario, in order to compute Eq. (\ref{eq:economy}), one is required to compute \(N_t\) for \(t \in [1, \dots, T]\). Hence, to obtain the number of removed plants, one can compute 
\begin{equation}
    \sum_{t=1}^{T} \frac{dR(t)}{dt}.
\label{eq:condition}
\end{equation}
Now, using the next generation matrix (NGM) method \cite{ngm}, we can bound the value of \(I(t)\) for each \(t \in [0, \dots, T]\) using the formula: \(I(t+1) \leq \beta I(t)\). By setting this condition to Eq. (\ref{eq:condition}) and using Eq. (\ref{eq:infection_rate}), one obtains:
\begin{equation}
    \sum_{t=1}^{T} \frac{dR(t)}{dt} = \sum_{t=1}^{T} \gamma I(t) = \sum_{t=1}^{T} \gamma \beta^{t-1} k \leq \gamma k \sum_{t=1}^{T} \frac{\beta_0^{t-1}}{r^{t-1}} = \frac{\gamma k \big ((\beta_0/r) ^T - 1 \big )}{(\beta_0/r) \big ((\beta_0/r) - 1 \big )}.
\label{eq:proof_results}
\end{equation}
From Eq. (\ref{eq:proof_results}), it is possible to obtain a close form for \[N_t \equiv N - R(t) = N - \Big ( \gamma k \big ((\beta_0/r) ^T - 1 \big ) \Big) / \Big ( (\beta_0/r) \big ((\beta_0/r) - 1 \big ) \Big ).\]
As a result, Eq. (\ref{eq:optimization}) can be rewritten as the following optimization tasks, solving for the worst-case scenario. For the conditions of either \(d_x > W\) or \(d_y > H\) no plants are seeded in the field which results in the target function to return 0. As a result, the problem takes the standard form of an optimization problem to \(N\):
\begin{equation}
    \begin{array}{l}
        \max_{d_x, d_y} \sum_{t=1}^T O(N_t)  \\
        \text{s.t.} \\
        d_x \leq W \\
        d_y \leq H \\
        d_x, d_y \geq 0
    \end{array}
\end{equation}
Assuming a decreased spatial step of size \(\delta \in \mathbb{R}^+\), one can solve this optimization task using brute force or using Monte-Carlo for good approximation with less computational burden \cite{optimization}. 

\section{Numerical Analysis}
\label{sec:results}
Using the proposed model and its implementation as a computer simulation, we investigate several scenarios of interest for the \textit{Rhizoctonia solani} pathogen that influences potatoes. The parameters used in the model calculations (if not stated otherwise) are presented in Table \ref{table:models_parameters}. The values are taken from the United States farming market price and constantly change over time. First, we show the average pandemic spread and economic profit for small, medium, and large fields. Subsequently, a sensitivity analysis of the pathogen and economic properties on the pandemic spread and economic profit are investigated. Finally, a comparison of default and optimal seeding strategies for random and worst-case initial conditions is analyzed. In order to capture the pandemic spread, we used the basic reproduction number (\(R_0\)) over time which is defined as \(R_0(t) := \frac{I(t+1)-I(t)}{R(t+1)-R(t)}\) for \(R(t+1) > R(t)\) and \(R_0(t) := I(t+1)-I(t)\), otherwise \cite{teddy_first}. In a complementary manner, the average basic reproduction number, \(E[R_0]\), is defined as follows: \(E[R_0] := \frac{1}{T}\sum_{t=1}^{T} R_0(t)\).

\begin{table}[!ht]
\center 
\begin{tabular}{p{0.07\textwidth}p{0.7\textwidth}p{0.1\textwidth}}
\hline
Symbol &  Parameter Definition & Value \\ \hline
\(S(t)\) & Susceptible plants  at time \(t\) [\(1\)] &   \\ 
\(I(t)\) & Infected plants at time \(t\) [\(1\)] &   \\ 
\(R(t)\) & Removed plants  at time \(t\) [\(1\)] &   \\ 
\(N_t\) & Number of plants in the population at time \(t\) [\(1\)] &  \\
\(r\) & The minimal distance of the maximum distance between an susceptible and infected plant at \(t=1\) [\(m\)] &  \\ \hline
 \(N\) & Default plant population size [\(plants\)] & \(25000\) \\ 
 \(d_x\) & Default x-axis distance between plants [\(m\)] & \(0.2\) \\ 
 \(d_y\) & Default y-axis distance between plants [\(m\)] & \(0.2\) \\
 \(W\) & The fields width  [\(m\)] & \(100\) \\
 \(H\) & The fields height  [\(m\)] & \(100\) \\   
 \(T\) & The duration of a session [\(plants\)]  & \(3\) \\ 
\(\beta_0\) & Basic infection probability [\(t^{-1} plants^{-1}\)] & 0.003  \\
\(\gamma\) & Removed rate [\(t^{-1}\)] & \(1/42\) \\  \(k\) & The initial number of infected plants at the beginning of the pandemic  [\(plants\)] & \(3\) \\  
\(a_1^s\) & Average cost per plant for seeding [\(\$\)] & 0.01\$ \\
\(a_2^s\) & Average overhead for seeding [\(\$\)] & 0.14N\$ \\  
\(a_1^g\) & Average cost per plant for growing per day [\(\$\)] & 0.033\$ \\
\(a_2^g\) & Average overhead for growing per day [\(\$\)] & 0.019N\$ \\  
\(a_1^h\) & Average cost per plant for harvesting [\(\$\)] & 0.06\$ \\
\(a_2^h\) & Average overhead for harvesting [\(\$\)] & 0.11N\$ \\  
\(\psi_1\) & Average cost per plant for selling [\(\$\)] & 5.32\$ \\
\(\psi_2\) & Average discount for selling plants [\(\$\)] & 1.71\$ \\  \hline
\end{tabular}
\caption{The model's parameters notations, descriptions, and values. }
\label{table:models_parameters}
\end{table}

\subsection{Baseline}
\begin{figure}[!ht]
    \centering
    \includegraphics[width=0.99\textwidth]{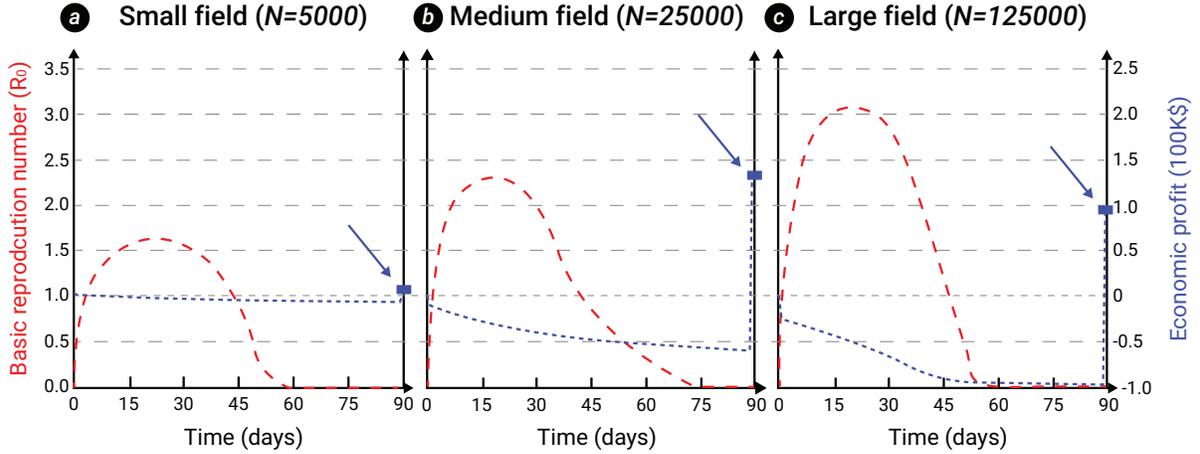}
\caption{The basic reproduction number (dashed, red) and economic profit (dotted, blue) over time, divided into (a) small, (b) medium, and (c) large size fields. The results are shown as an average of \(n=1000\) random instances with the parameter values from Table \ref{table:models_parameters}.}
    \label{fig:baseline}
\end{figure}

Fig. \ref{fig:baseline} shows the basic reproduction number and economic profit over time, divided into small (\(N=5000\)), medium (\(N=25000\)), and large (\(N=125000\)) fields with the same size. All fields are of the same size. The results show the average of \(n=1000\) simulations for each field that differ from each other by the position of initially infected plants. The overall infection rate over time increases as the field size increases and the distance between plants decreases. As a direct outcome, the economic profit at the end of the session is not monotonically increasing as expected (assuming it is profitable to grow a single plant) since a more significant portion of the plants is removed.

\subsection{Sensitivity analysis}
\begin{figure}[!ht]
    \centering
    \includegraphics[width=0.99\textwidth]{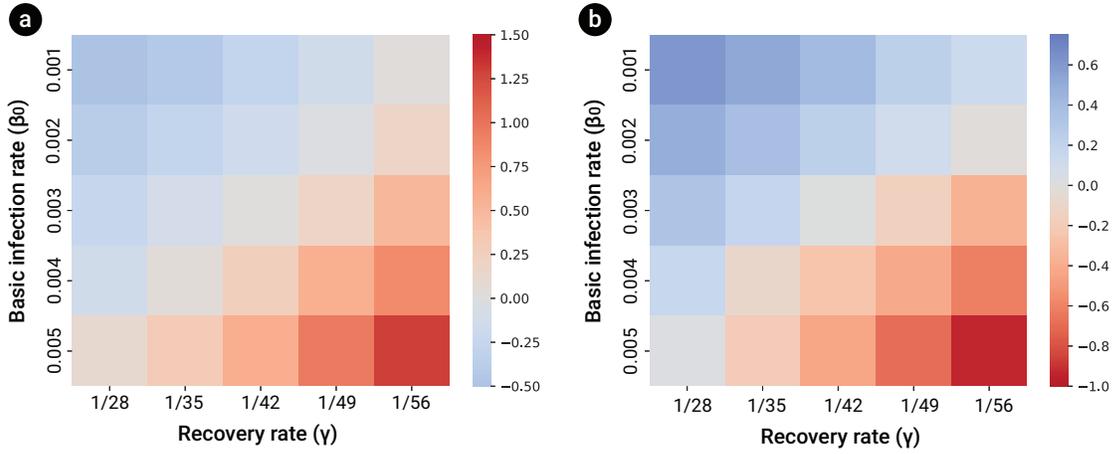}
\caption{Sensitivity analysis of the (a) average basic reproduction number and (b) economic profit as a function of the basic infection rate and recovery rate, relative to the baseline configuration - \(\beta_0 = 0.003, \gamma = 1/42\). The results are shown as an average of \(n=1000\) random instances with the parameter values from Table \ref{table:models_parameters}. The reader should note the different color scales in panels (a) and (b).}
    \label{fig:sens_pandemic}
\end{figure}

In order to capture the influence of the epidemiological parameters associated with the pathogen, \(\beta_0\) and \(\gamma\), a sensitivity analysis of the pandemic spread as the average basic reproduction number (a) and economic profit (b) are shown in Fig. \ref{fig:sens_pandemic}. The results are shown relative to the baseline configuration provided in Table \ref{table:models_parameters} where \(\beta_0 = 0.003\) and \(\gamma = 1/42\). We fitted the numerical data with an analytical two-dimensional linear function using the least mean square (LMS) method \cite{leastSquares}, obtaining:
\begin{equation}
    E[R_0] = 226.61 \beta_0 - 42.88 \gamma + 0.53 \wedge E(T) = -213.18 \beta_0 + 37.27 \gamma - 0.32,  
    \label{eq:sens_pandemic_fitted}
\end{equation}
with a coefficient of determination (\(R^2\)) of \(0.896\) and \(0.954\), respectively.

\begin{figure}[!ht]
    \centering
    \includegraphics[width=0.99\textwidth]{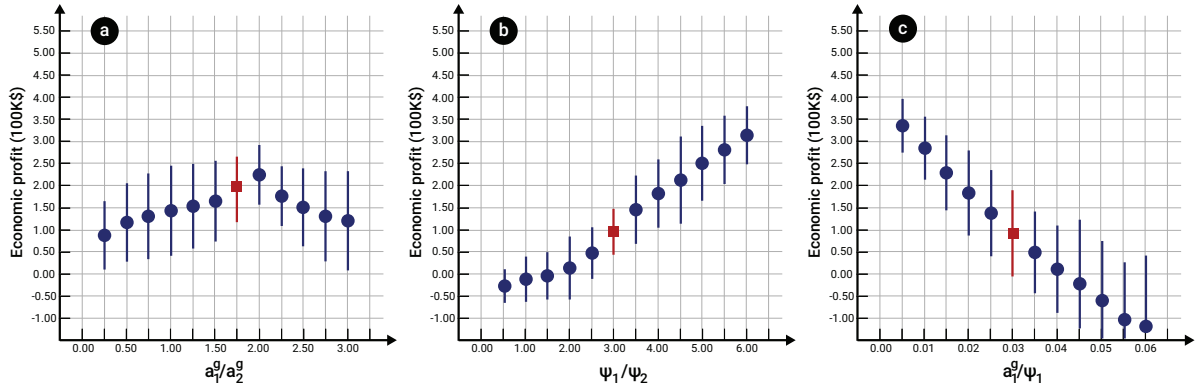}
\caption{Sensitivity analysis of the economic profit as a function of several economic ratios - (a) \(a_1^g/a_2^g\), (b) \(\psi_1/\psi_2\), and (c) \(a_1^g/\psi_1\). The results are shown as an average \(\pm\) standard deviation of \(n=1000\) random instances with the parameter values from Table \ref{table:models_parameters}. The square (red) value indicates the baseline configuration shown in Table \ref{table:models_parameters}.}
    \label{fig:sens_economy}
\end{figure}

In addition, we measured the influence of three economic ratios \(a_1^g/a_2^g\), \(\psi_1/\psi_2\), and \(a_1^g/\psi_1\) on the economic profit as shown in Fig. \ref{fig:sens_economy}. The results are shown as mean \(\pm\) standard deviation of \(n=1000\) instances. The square (red) value indicates the baseline configuration shown in Table \ref{table:models_parameters}. One can notice that \(a_1^g/a_2^g\) obtains an optimum around \(2\). The economic profit is monotonically increasing as a function of \(\psi_1/\psi_2\). Oppositely, the economic profit is monotonically decreasing as a function of \(a_1^g/\psi_1\).

\subsection{Optimal seeding strategy}
\begin{figure}[!ht]
    \centering
    \includegraphics[width=0.99\textwidth]{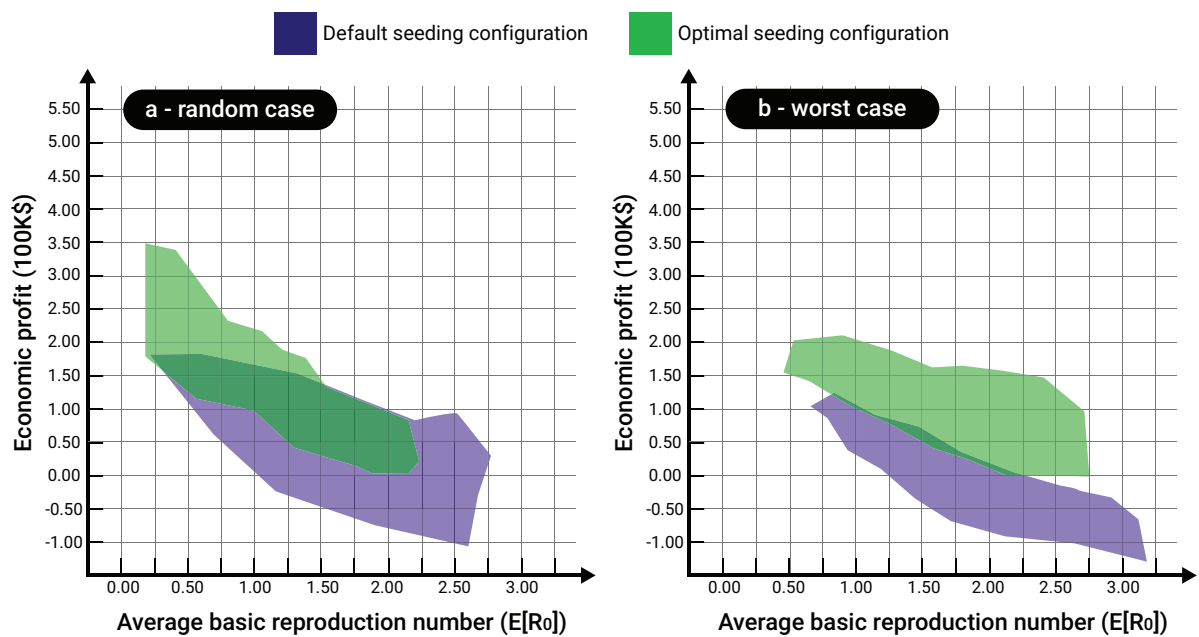}
\caption{The average pandemic spread and economic profit for the (a) random case and (b) worst-case optimal seeding strategy. The results are based on \(n=10000\) random instances with the parameter values from Table \ref{table:models_parameters} and \(W \in [50, 250], H \in [50, 250], \beta_0 \in [0.001, 0.005], \gamma \in [1/65, 1/21]\).}
    \label{fig:optimal}
\end{figure}

Following the algorithm for the optimal seeding strategy (see Section \ref{sec:worse_case}), we computed the difference between the default (\(d_x = d_y = 0.2\)) and optimal seeding strategy for random and worst case of initial condition. The results are summarized in Fig. \ref{fig:optimal} such that the results obtained from \(n=10000\) random instances where the field size (\(W, H\)) and pathogen properties (\(\beta_0, \gamma\)) are different for each instance. The blue and green area indicates the convex hull obtained from the dots using the Graham scan algorithm \cite{graham}. 

In order to capture the underline functional dynamics generating the economic profit from the economic and epidemiological parameters, we utilized the \textit{SciMED} symbolic regression tool \cite{scimed}. Symbolic regression involves conducting regression analysis by exploring the realm of mathematical expressions to discover the most suitable model for a provided dataset, prioritizing both accuracy and simplicity. We obtained the following equation:
\begin{equation}
    E(T) = (\psi_1 - a_1^gT \frac{\beta_0 r^2}{\gamma} + a_2^gT) N,
    \label{eq:scimed}
\end{equation}
with a coefficient of determination \(R^2 = 0.741\) for the random case with the optimal seeding strategy.

\section{Discussion}
\label{sec:discussion}
In this study, we have developed a mathematical model to establish the economic profit from a field of plants during a botanical pandemic. The proposed model establishes the connection between a spatio-temporal SIR-based epidemiological model and a non-linear economic production model. Based on this model, we propose an algorithm to find the economically optimal seeding strategy given a known pathogen and economic properties.

Considering \textit{Rhizoctonia solani} for potatoes as a representative example, we implemented an agent-based simulation. We studied the influence of several environmental, epidemiological, and economic properties of the pandemic spread and economic profit. In particular, we demonstrate the potential benefits of the proposed seeding strategy for both a random and worst-case pandemic scenario, regardless of the pathogen's properties and field size. 

The baseline analysis aims to indicate a connection between the plant population size and the pandemic spread. On average, as the plant population size increases, the infection rate, reflected by the mean basic reproduction number (\(R_0\)), increases as well, as shown in Fig. \ref{fig:baseline}. Moreover, the economic profit, at the end of the season (\(T\)), first increases and then decreases, demonstrating a parabolic-like behavior. These results agree with multiple spatio-temporal epidemiological models, and historical data in human and animal pandemics \cite{spatial_2,graph_3}.

Following the same path, an analysis of the pandemic spread and economic profit as a function of the basic infection rate (\(\beta_0\)) and recovery rate (\(\gamma\)) is conducted as shown in Fig. \ref{fig:sens_pandemic}. Unsurprisingly, as the basic infection rate increases and the recovery rate decreases, the pandemic spread increases, and the economic profit decreases. Moreover, from Eq. (\ref{eq:sens_pandemic_fitted}), one can see that the basic infection rate has \(~6\) times more influence on the decrease in economic profit compared to the recovery rate. Thus, when designing pandemic intervention policies, researchers and developers might prefer to focus on the pandemic intervention policies or the development of plants that are more resilient to an infection to reduce the pandemic spread. 

In an interconnected way, the sensitivity analysis of the economic profit as a function of the ratio between growing overhead and per plant cost, the ratio between the average plant selling price and overall reduced price, and the ratio between average growing cost per plant to the selling cost is studied in Fig. \ref{fig:sens_economy}. Fig. \ref{fig:sens_economy}a reveals that as the ratio between the per plant growing cost and the growing overhead cost increases, it reaches an optimum and then sharply decreases. However, for the two other quantities, the dynamics are monotonic, agreeing with previous economic models \cite{atkeson2021economic,graph_based_3}.

Most importantly, Fig. \ref{fig:optimal} reveals that the optimal seeding strategy indeed increases the average (and in general) economic profit while decreasing the pandemic spread for both the random and worst-case scenarios. Specifically, the improvement for the random case is statistically significantly better than for the worst-case scenario (\(p < 0.005\)), obtained using an ANOVA test. Nevertheless, for both the random and worst case, the optimal seeding configuration is statistically outperforming the default seeding configuration (\(p < 0.001\)), obtained using a paired two-sided T-test. Interestingly, while the relationship between the average basic reproduction number and the economic profit for both cases is complex and non-linear, the figure reveals a somewhat linear decreasing trend. Moreover, one can notice that the improvement of using the optimal seeding configuration compared to the default seeding configuration in the random case has less impact compared to the worst case. 

Altogether, field owners and policymakers can adopt the seeding strategy at the beginning of a new session, assuming they are familiar with the possible pathogen that might cause a pandemic in their field, given a crop they wish to grow. Moreover, the proposed model provides a tool to approximate the economic influence such a pandemic might have on a specific situation. Our model and simulator are published as open source\footnote{The source code is available at \url{https://github.com/teddy4445/pandemic_in_the_field}}, so other researchers and policymakers can replicate and extend our study for their needs.  

This study has several limitations, which provide opportunities for future research. First, the proposed model assumes that the plant population is homogeneous in its epidemiological and economic properties, an assumption known to be false \cite{discussion_support_multi_type,stochastic_example_1}. As such, by introducing unique epidemiological and economic properties for each individual plant, such as recovery rate and amount of profit from selling, one would obtain a more accurate economic profit prediction. In a complementary manner, the proposed work focused on the seeding strategy alone, ignoring current and novel possible monitoring and intervention policies strategies that can control a pandemic spread and therefore increase the economic profit \cite{limits_1,limits_2}. In addition, as each plant type is susceptible to multiple pathogens, an extension of the proposed model for the case of a multi-strain pandemic can be of great interest \cite{multi_strain_1,multi_strain_2,multi_strain_5,multi_strain_3,multi_strain_4}. Finally, the proposed analysis was conducted for a single pathogen, as shown in Table \ref{table:models_parameters}, which limits our current understanding of the model's generalization capabilities. Future work can explore the proposed model for more pathogens. Moreover, in this work, we assume a two-dimensional field. However, many fields take advantage of three-dimensional phenomena. A possible extension of this work is to include a three-dimensional surface.

\nolinenumbers

\section*{Declarations}

\subsection*{Funding}
This research did not receive any specific grant from funding agencies in the public, commercial, or not-for-profit sectors.

\subsection*{Conflicts of interest/Competing interests}
None.

\subsection*{Data availability}
The data that has been used is presented in the manuscript with the relevant sources.

\subsection*{Code availability}
The code and materials are available upon request. 
 
\bibliography{biblio}

\begin{thebibliography}{10}

\bibitem{intro}
R.~A.~C. Jones.
\newblock Global plant virus disease pandemics and epidemics.
\newblock {\em Plants}, 10(2), 2021.

\bibitem{intro_2}
J.~C. Zadoks.
\newblock Twenty-five years of botanical epidemiology.
\newblock {\em Phil. Trans. R. Soc. Lond. B}, 321:377–387, 1988.

\bibitem{bp_1}
J.~M. Thresh.
\newblock {\em The impact of plant virus diseases in developing countries}.
\newblock Virus and Virus-like Diseases of Major Crops in Developing Countries,
  2004.

\bibitem{bp_2}
M.J. Jeger and J.M. Thresh.
\newblock Modelling reinfection of replanted cocoa by swollen shoot virus in
  pandemically diseased areas.
\newblock {\em J. Appl. Ecol.}, 30:187–196, 1993.

\bibitem{bp_3}
G.~W. Otim-Nape and J.~M. Thresh.
\newblock {\em The current pandemic of cassava mosaic virus disease in Uganda}.
\newblock The Epidemiology of Plant Diseases, 1998.

\bibitem{s_reviewer_1}
D.~J. Bailey, W.~Otten, and C.~A. Gilligan.
\newblock Saprotrophic invasion by the soil-borne fungal plant pathogen
  rhizoctonia solani and percolation thresholds.
\newblock {\em New Phytologist}, 146(3):535--544, 2000.

\bibitem{s_reviewer_2}
S.~Poggi, F.~M. Neri, V.~Deytieux, A.~Bates, W.~Otten, C.~A. Gilligan, and
  D.~J. Bailey.
\newblock Percolation-based risk index for pathogen invasion: Application to
  soilborne disease in propagation systems.
\newblock {\em Phytopathology}, 103(10):1012--1019, 2013.

\bibitem{related_1}
C.~A. Gilligan, S.~Gubbins, and S.~A. Simons.
\newblock Analysis and fitting of an {SIR} model with host response to
  infection load for a plant disease.
\newblock {\em Philosophical Transactions of the Royal Society of London.
  Series B: Biological Sciences}, 352(1351):353--364, 1997.

\bibitem{intro_3}
L.V. Madden.
\newblock Botanical epidemiology: Some key advances and its continuing role in
  disease management.
\newblock {\em Eur J Plant Pathol}, 115:3–23, 2006.

\bibitem{pandemic_economy}
S.~Galiani.
\newblock On the application of the window of opportunity and complex network
  to risk analysis of process plants operations during a pandemic.
\newblock {\em Journal of Economic Behavior \& Organization}, 193:269--275,
  2022.

\bibitem{intro_money}
S.~K. Sastry and T.~A. Zitter.
\newblock {\em Management of virus and viroid diseases of crops in the
  tropics}, volume~2.
\newblock Plant Virus and Viroid Diseases in the Tropics, Epidemiology and
  Management, 2014.

\bibitem{intro_economy_1}
P.~K. Anderson, A.~A. Cunningham, N.~G. Patel, F.~J. Morales, P.~R. Epstein,
  and P.~Daszak.
\newblock Emerging infectious diseases of plants: Pathogen pollution, climate
  change and agrotechnology drivers.
\newblock {\em Trends Ecol. Evol.}, 19:535--544, 2004.

\bibitem{intro_economy_2}
J.~M. Thresh.
\newblock The origins and epidemiology of some important plant virus diseases.
\newblock {\em Appl. Biol.}, 5:1--65, 1980.

\bibitem{intro_economy_3}
J.~M. Thresh.
\newblock Cropping practices and virus spread.
\newblock {\em Annu. Rev. Phytopathol.}, 20:193--218, 1982.

\bibitem{intro_related_1}
P.~Kampmeijer and J.~C. Zadoks.
\newblock Asimulatoroffoci and epidemics in mixtures, multilines, and mosaics
  of resistant and susceptible plants.
\newblock {\em EPIMUL}, page~50, 1997.

\bibitem{intro_related_3}
L.V. Madden and F.~Van~den Bosch.
\newblock A population-dynamics approach to assess the threat of plant
  pathogens as biological weapons against annual crops.
\newblock {\em BioScience}, 52:65--74, 2002.

\bibitem{intro_related_4}
C.~A. Gilligan and S.~Gubbins.
\newblock Analysis and fitting of an sir model with host response to infection
  load for a plant disease.
\newblock {\em Philos. Trans. R. Soc. B Biol. Sci.}, 352:353–364, 1997.

\bibitem{first_sir_paper}
W.~O. Kermack and A.~G. McKendrick.
\newblock A contribution to the mathematical theory of epidemics.
\newblock {\em Proceedings of the Royal Society}, 115:700–721, 1927.

\bibitem{related_2}
A.~Kawaguchi, S.~Kitabayashi, K.~Inoue, and K.~Tanina.
\newblock An hld model for tomato bacterial canker focusing on epidemics of the
  pathogen due to cutting by infected scissors.
\newblock {\em Plants}, 11(17), 2022.

\bibitem{related_3}
N.~J. Cunniffe and C.~A. Gilligan.
\newblock Invasion, persistence and control inepidemic models for plant
  pathogens: the effect of host demography.
\newblock {\em Journal of the Royal Society Interface}, 7:439--451, 2010.

\bibitem{teddy_ariel}
T.~Lazebnik and A.~Alexi.
\newblock Comparison of pandemic intervention policies in several building
  types using heterogeneous population model.
\newblock {\em Communications in Nonlinear Science and Numerical Simulation},
  107(4):106176, 2022.

\bibitem{abs_1}
D.~Chumachenko, V.~Dobriak, M.~Mazorchuk, I.~Meniailov, and K.~Bazilevych.
\newblock On agent-based approach to influenza and acute respiratory virus
  infection simulation.
\newblock In {\em 2018 14th International Conference on Advanced Trends in
  Radioelecrtronics, Telecommunications and Computer Engineering (TCSET)},
  pages 192--195, 2018.

\bibitem{abs_2}
J.~D. Priest, A.~Kishore, L.~Machi, C.~J. Kuhlman, D.~Machi, and S.~S. Ravi.
\newblock Csonnet: An agent-based modeling software system for discrete time
  simulation.
\newblock In {\em 2021 Winter Simulation Conference (WSC)}, pages 1--12, 2021.

\bibitem{abs_3}
K.~M. Carley, D.~B. Fridsma, E.~Casman, A.~Yahja, N.~Altman, Li-Chiou C.,
  B.~Kaminsky, and D.~Nave.
\newblock Biowar: scalable agent-based model of bioattacks.
\newblock {\em IEEE Transactions on Systems, Man, and Cybernetics - Part A:
  Systems and Humans}, 36(2):252--265, 2006.

\bibitem{spatial}
N.~Cressie.
\newblock {\em Statistics for spatial data}.
\newblock John Wiley \& Sons, 1991.

\bibitem{price_1}
D.~Levy, S.~Dutta, M.~Bergen, and R.~Venable.
\newblock Price adjustment at multiproduct retailers.
\newblock {\em Managerial and decision economics}, 19(2):81--120, 1998.

\bibitem{price_2}
P.~J. Klenow and B.~A. Malin.
\newblock Microeconomic evidence on price-setting.
\newblock In {\em Handbook of monetary economics}, volume~3, pages 231--284.
  Elsevier, 2010.

\bibitem{economic_1}
M.~G. Dekimpe and D.~M. Hanssens.
\newblock Advertising response models.
\newblock In {\em The Sage Handbook of Advertising}, pages 247--263. Sage
  Publications, 2007.

\bibitem{economic_2}
J.~L. Simon.
\newblock {\em Issues in the Economics of Advertising}.
\newblock University of Illinois Press, 1970.

\bibitem{economic_3}
J.~L. Simon and J.~Arndt.
\newblock The shape of the advertising response function.
\newblock {\em Journal of Advertising Research}, 20(4), 1980.

\bibitem{agent_based_1}
J.~D. Priest, A.~Kishore, L.~Machi, C.~J. Kuhlman, D.~Machi, and S.~S. Ravi.
\newblock Csonnet: An agent-based modeling software system for discrete time
  simulation.
\newblock In {\em 2021 Winter Simulation Conference (WSC)}, pages 1--12, 2021.

\bibitem{agent_based_2}
L.~Tesfatsion.
\newblock Agent-based computational economics: Growing economies from the
  bottom up.
\newblock {\em Artificial Life}, 8(1), 2002.

\bibitem{teddy_pandemic_management}
T.~Lazebnik, S.~Bunimovich-Mendrazitsky, and L.~Shami.
\newblock Pandemic management by a spatio–temporal mathematical model.
\newblock {\em International Journal of Nonlinear Sciences and Numerical
  Simulation}, 107(4):106176, 2021.

\bibitem{fsm}
V.~S. Alagar and K.~Periyasamy.
\newblock {\em Extended Finite State Machine}, pages 105--128.
\newblock Springer London, 2011.

\bibitem{k_center}
J.~Chuzhoy, S.~Guha, E.~Halperin, S.~Khanna, G.~Kortsarz, R.~Krauthgamer, and
  J.~Naor.
\newblock Asymmetric k-center is log* n-hard to approximate.
\newblock {\em J. ACM}, 52(4):538–551, 2005.

\bibitem{ngm}
O.~Diekmann, J.~A. Heesterbeek, and M.~G. Roberts.
\newblock The construction of next-generation matrices for compartmental
  epidemic models.
\newblock {\em Journal of the Royal Society}, 7(47):873–885, 2010.

\bibitem{optimization}
E.~Zio.
\newblock {\em Monte Carlo Simulation: The Method}, pages 19--58.
\newblock Springer London, London, 2013.

\bibitem{teddy_first}
T.~Lazebnik and S.~Bunimovich-Mendrazitsky.
\newblock The signature features of covid-19 pandemic in a hybrid mathematical
  model - implications for optimal work-school lockdown policy.
\newblock {\em Advanced theory and simulations}, 2021.

\bibitem{leastSquares}
A.~Bjorck.
\newblock Numerical methods for least squares problems.
\newblock {\em Society for Industrial and Applied Mathematics}, 5:497--513,
  1996.

\bibitem{graham}
R.~L. Graham.
\newblock An efficient algorithm for determining the convex hull of a finite
  planar set.
\newblock {\em Information Processing Letters}, 1(4):132–133, 1972.

\bibitem{scimed}
L.~S. Keren, A.~Liberzon, and T.~Lazebnik.
\newblock A computational framework for physics-informed symbolic regression
  with straightforward integration of domain knowledge.
\newblock {\em Scientific Reports}, 13:1249, 2023.

\bibitem{spatial_2}
M.~J. Keeling and K.~T.~D. Eames.
\newblock Networks and epidemic models.
\newblock {\em Journal of The Royal Society Interface}, 2:295--307, 2005.

\bibitem{graph_3}
C.~Moore and M.~E.~J. Newman.
\newblock Epidemics and percolation in small-world networks.
\newblock {\em Phys. Rev. E}, 61:5678--5682, 2000.

\bibitem{atkeson2021economic}
A.~Atkeson, M.~C. Droste, M.~J. Mina, and J.~Stock.
\newblock Economic benefits of covid-19 screening tests with a vaccine rollout.
\newblock {\em medRxiv}, 2021.

\bibitem{graph_based_3}
M.~Bognanni, H.~Doug, D.~Kolliner, and K.~Mitman.
\newblock Economics and epidemics: Evidence from an estimated spatial econ-sir
  model.
\newblock {\em Finance and Economics Discussion Series 2020-091. Washington:
  Board of Governors of the Federal Reserve System}, 2020.

\bibitem{discussion_support_multi_type}
H.~Sun, H.~Wang, M.~Yang, and G.~Reniers.
\newblock On the application of the window of opportunity and complex network
  to risk analysis of process plants operations during a pandemic.
\newblock {\em Journal of Loss Prevention in the Process Industries},
  68:104322, 2020.

\bibitem{stochastic_example_1}
C-H. Li, C-C. Tsai, and S-Y. Yang.
\newblock {Analysis of epidemic spreading of an SIRS model in complex
  heterogeneous networks}.
\newblock {\em Communications in Nonlinear Science and Numerical Simulation},
  19(4):1042--1054, 2014.

\bibitem{limits_1}
J.~B. Ristaino, P.~K. Anderson, D.~P. Bebber, K.~A. Brauman, N.~J. Cunniffe,
  N.~V. Fedoroff, C.~Finegold, K.~A. Garrett, C.~A. Gilligan, C.~M. Jones,
  M.~D. Martin, G.~K. MacDonald, P.~Neenan, A.~Records, D.~G. Schmale,
  L.~Tateosian, and Q.~Wei.
\newblock The persistent threat of emerging plant disease pandemics to global
  food security.
\newblock {\em Proceedings of the National Academy of Sciences},
  118(23):e2022239118, 2021.

\bibitem{limits_2}
J.~P. Legg, R.~Shirima, L.~S. Tajebe, D.~Guastella, S.~Boniface, S.~Jeremiah,
  E.~Nsami, P.~Chikoti, and C.~Rapisarda.
\newblock Biology and management of bemisia whitefly vectors of cassava virus
  pandemics in africa.
\newblock {\em Pest Management Science}, 70(10):1446--1453, 2014.

\bibitem{multi_strain_1}
I.~Gordo, M.~G.~M. Gomes, D.~G. Reis, and P.~R.~A. Campos.
\newblock Genetic diversity in the {SIR} model of pathogen evolution.
\newblock {\em Plos One}, 4(3):e4876, 2009.

\bibitem{multi_strain_2}
O.~Khyar and K.~Allali.
\newblock Global dynamics of a multi-strain seir epidemic model with general
  incidence rates: application to covid-19 pandemic.
\newblock {\em Nonlinear Dynamics}, 102:489--509, 2020.

\bibitem{multi_strain_5}
T.~Lazebnik and S.~Bunimovich-Mendrazitsky.
\newblock Generic approach for mathematical model of multi-strain pandemics.
\newblock {\em Plos One}, 17(4):e0260683, 2022.

\bibitem{multi_strain_3}
P.~Minayev and N.~Ferguson.
\newblock Improving the realism of deterministic multi-strain models:
  implications for modelling influenza a.
\newblock {\em Journal of the Royal Society Interface}, 2008.

\bibitem{multi_strain_4}
Y-X. Dang, X-Z. Li, and M.~Martcheva.
\newblock Competitive exclusion in a multi-strain immuno-epidemiological
  influenza model with environmental transmission.
\newblock {\em Journal of Biological Dynamics}, 10(1), 2016.

\end{thebibliography}
\bibliographystyle{unsrt}

\end{document}